# Designing an Inclusive and Engaging Hybrid Event: Experiences from CHIWORK

**ANDREW L. KUN**
University of New Hampshire, andrew.kun@unh.edu

**ORIT SHAER**
Wellesley College, oshaer@wellesley.edu

Conferences are a key place for training and education. Attendees learn about the state of the art of their field, about relevant methods, and acquire networking skills that can support their work.

In the world that was changed by the COVID-19 pandemic [1], we often need to organize hybrid training and education events, including conferences, with both in-person and remote attendees [2]. Both organizers and attendees are eager for events that are productive, safe, and that bring together a diverse group of our colleagues, across multiple fields of study, multiple countries, as well as with different capacities to travel and attend in-person and remote meetings. However, the best practices for such hybrid events are still under development. In this document we hope to contribute to this development of best practices: we report on our experiences in organizing a small, hybrid conference, and provide the lessons we learned through this process.

## 1 ORGANIZING A HYBRID EVENT

Our focus is the first CHIWORK annual meeting, which ran as a hybrid event June 8-9, 2022, in Durham, NH, USA. A total of 115 participants registered for the annual meeting, 27 for in-person attendance and 88 for online attendance. During the two days of the conference each of our sessions was attended by approximately 20-25 in-person participants, and 20-25 online participants. We received 30 full paper submissions of which 18 were accepted.

The meeting was conducted during the COVID-19 pandemic, and so we prioritized safety. Our approach was to create a hybrid event, which allows participants the flexibility to attend CHIWORK in the way that suited their safety needs. We also instituted active safety measures at the meeting: we had daily, self-administered COVID tests, and we mandated N95 or KN95 masks for all. We also encouraged participants to eat outside, and to not eat indoors.

For supporting diversity and inclusivity, we again believed that a hybrid event was the right approach. This allows for a wide range of voices to be heard. The first questions we received when we announced the call-for-papers were: "do we have to travel to present?" We would have lost many participants if we had mandated in-person presentations/attendance. This would have harmed the diversity of perspectives and depth of discussion during CHIWORK.

We recognized that inclusive and engaging hybrid experiences are difficult to create, and some might argue even impossible to achieve (cf. [3]). We had no doubt that the in-person and online experiences would not be the same. For in-person attendees, it would be possible to be present, to go to dinners, to chat in the hallways, and to hear applause. Online participants would miss these parts of the event. So, we dedicated much thought to how to integrate online and in-person participants. In the following we describe our design decisions.

### 1.1 Presentation format

We asked authors, both online and in-person, how they would like to present their work. They had two options: they could present their work with a standard slide presentation, or they could have a conversation about their paper with a session chair (Figure 1). For conversations, the session chair would share a file with the presenter outlining the timing of the conversation and providing several questions in advance. We asked presenters to outline their responses in the shared file. The questions and responses served as scaffolding for the presenter and session chair to engage in a lively discussion about the paper. It is this discussion that replaced the standard slide presentation and was intended to provide the basic information about the paper to the audience. After this planned part of the discussion, the session chair would turn to the audience for additional questions – just as they would do after a slide presentation.

We ended up with a variety of presentation formats: in-person conversations and slide shows, as well as remote conversations and slide shows. Adapting the presentation style to the desires of the authors made authors comfortable. At the same time, the variety of presentation styles helped online and in-person participants focus on the content. Also, the variety of formats within each session made the session feel more dynamic and less monotonous than webinar or online presentation sessions.

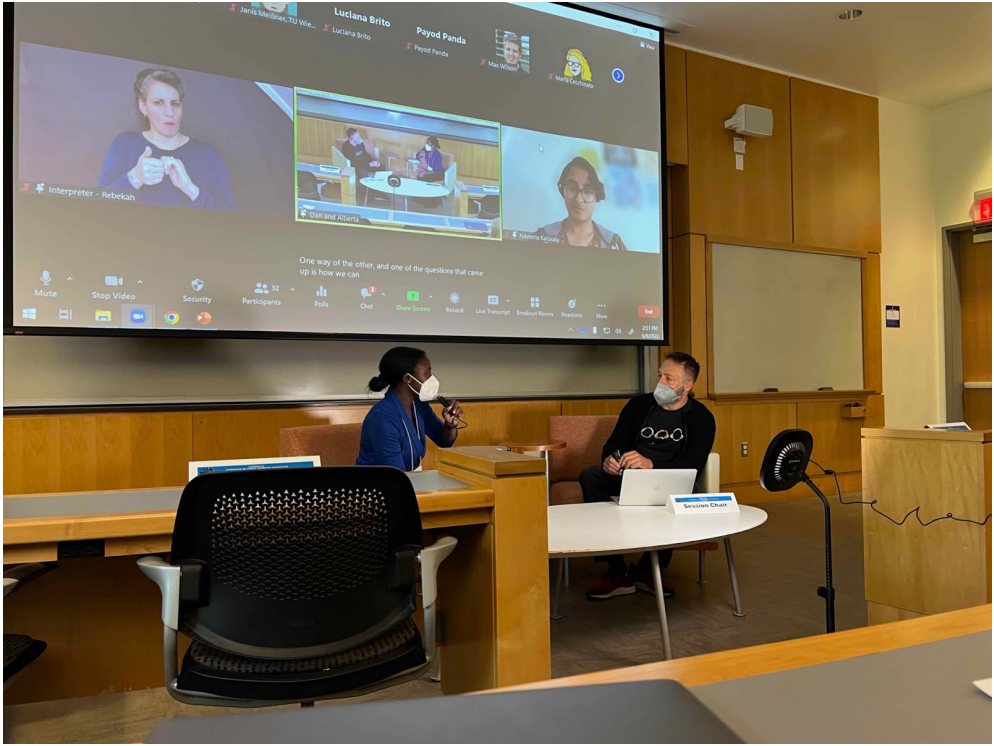

Figure 1: An in-person conversation at CHIWORK 2022. Alberta Ansah (left) talks to in-person session chair, Daniele Quercia. The screen shows the spotlighted windows: the sign language interpreter (left), the local view (middle), and the online chair, Naveena Karusala (right).

### 1.2 Session chairing

We wanted to make the hybrid experience relatable to both online and in-person participants. For this reason, we assigned two session chairs for each session: an online session chair and an in-person session chair. Online session chairs monitored questions from online participants, making sure their voices were heard. In-person chairs did the same for in-person participants. Online chairs presented remote speakers and moderated conversations with remote participants. In person session chairs moderated in-person presentations and conversations.

### 1.3 Digital setup

All papers sessions, and the two keynote talks were streamed via Zoom.

For slide presentations (with online and in-person presenters), we shared the slides on Zoom. For in-person conversations, we used Zoom to show the presenter and session chair, who were comfortably seated at the front of the meeting room.

We used the "spotlight" feature of Zoom to place key camera views in prominent and predictable places on remote participants' screens. Specifically, we spotlighted the virtual chair and the local chair. For in-person presentations and conversations, we spotlighted the main camera of the meeting room. For presentations we would focus the camera on the presenter who would stand by a podium. For conversations we would aim the camera at two armchairs placed in the front of the meeting room. For virtual presentations and conversations, we spotlighted the remote presenter. We also spotlighted sign language interpreters.

We had breakout rooms in Zoom where online participants could chat during breaks. This way they could use audio without having to worry that they are interfering with the in-person participants (simply speaking in the main Zoom meeting would be heard at the in-person venue).

We used Discord as the communication medium for online and in-person attendees. We did not want participants to use Zoom for chatting, since we expected that this would divide participants into the online group that uses Zoom, and the in-person group that does not use Zoom.

## 1.4 Physical setup

We had an excellent meeting room at the University of New Hampshire that was equipped with two cameras (front and back of the room), and with excellent audio (four mobile microphones). We added our own video cameras for better visibility of the in-person sign language interpreter and for in-person questions (Figure 2). We also added lighting so that in-person presenters, session chairs, and participants asking questions would be better seen on Zoom by remote participants (Figure 2).

We placed armchairs at the front of the room for in-person conversations (Figure 3). These armchairs conveyed a sense of calm and comfort that we hoped would put everyone at ease.

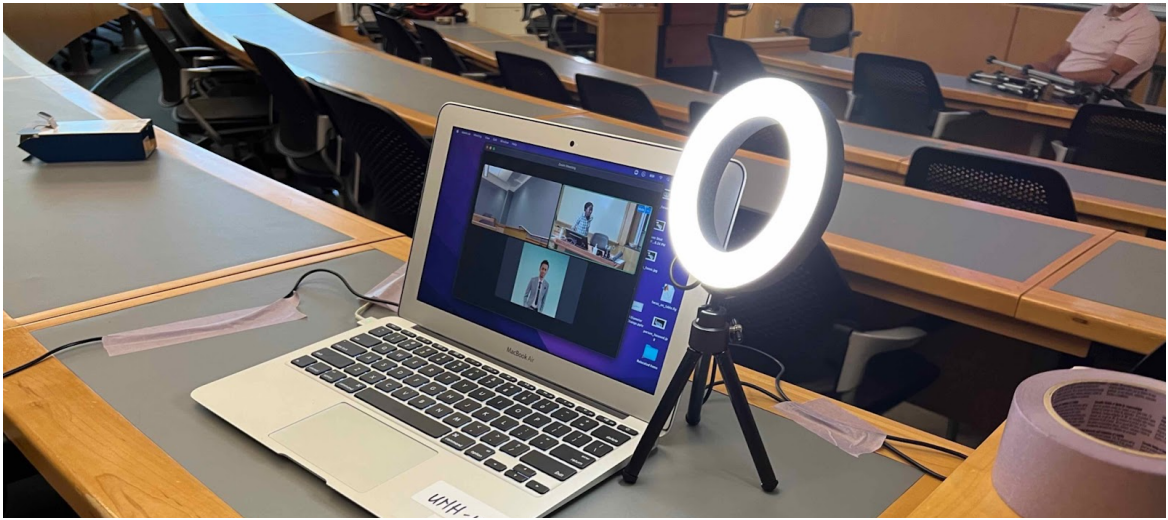

Figure 1: We set up laptops for in-person participants to ask questions. This way online participants could clearly see who is asking the question. We added lights to improve video quality.

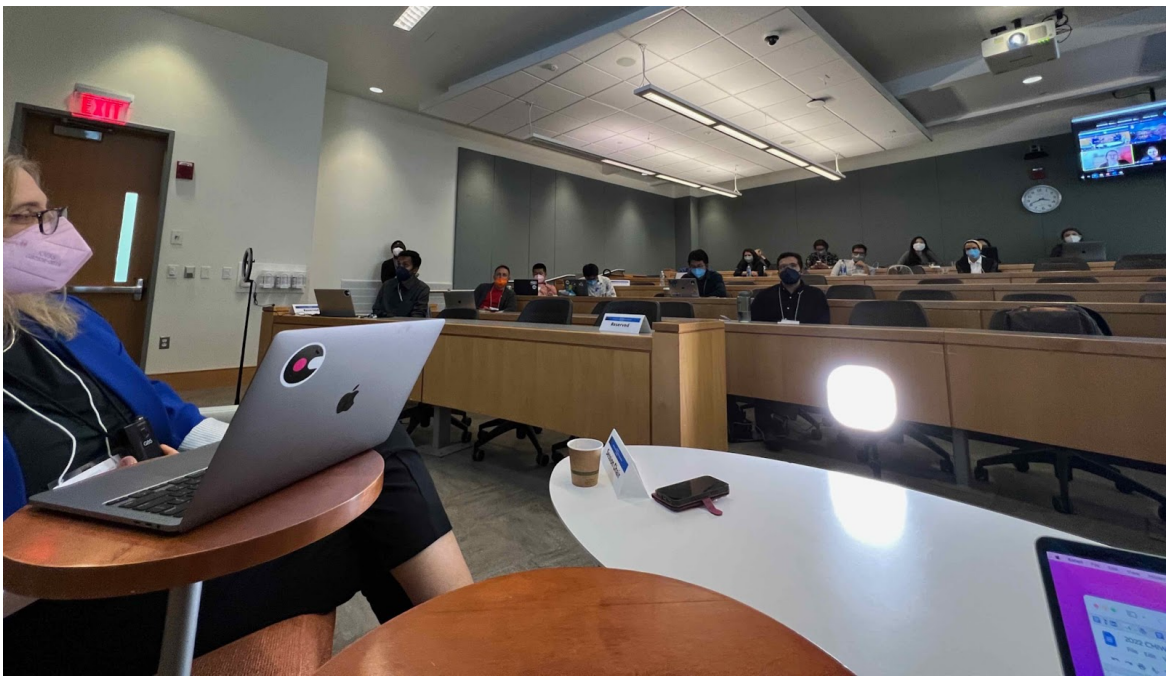

Figure 3: The authors' view sitting in the armchairs during the town hall, which was held at the end of the two-day meeting.

## 1.5 Accessibility

Accessibility was foremost on our minds as we designed the CHIWORK Annual Meeting.

First, we enabled captions in Zoom to support all online participants. Captions support participants with hearing impairment, and also those who might find it challenging to follow presentations and conversations in English.

We secured American sign-language interpreters for the conference. The interpreters worked online. And, for many of us, the CHIWORK Annual Meeting was our first opportunity to hear a presentation in which the presenters used sign language, which was instantly converted to spoken English by the sign-language interpreters.

Also in support of accessibility, we asked in-person participants to always use microphones so that they were clearly heard both at the venue and online. Also, we added cameras for in-person participants to use when asking questions – participants had to move to one of two laptops with cameras when asking a question.

## 1.6 Organization

All the actions above needed to be executed on a tight schedule. We needed to know, at all times: (1) who should be spotlighted in Zoom, (2) who should be taking questions from the audience, and (3) which camera should we turn on, the one showing the front of the in-person venue or the audience? To be able to execute all these actions at the right time, we created a minute-by-minute breakdown of actions for us to follow – we called this our "run sheet." The run sheet was in a shared spreadsheet, accessed by the organizers and student volunteers. When an action was taken care of, we would place a checkmark next to it, and move on. And of course, we practiced all these actions. We had 1-2 hour-long practice sessions almost every day for the two weeks leading up to the hybrid event. We used these practice sessions to perfect our plans, to become quick in executing them, and also to invite remote presenters to join us if they wanted to test out their setup for remote presentation.

## 2 EVALUATION

We evaluated the 2022 CHIWORK Annual Meeting in two ways – through a survey, and through feedback received at the town hall, which was held as the last session of the meeting. We discuss the feedback in the next two sections – overall, the feedback was encouraging, and it included insightful suggestions for improvements.

## 2.1 Survey

After the event, we sent out an anonymous survey, modeled on the survey used by Schloss Dagstuhl – Leibniz Center for Informatics[1]. We received 22 responses, with 15 responses indicating that they attended remotely, and 7 that they attended in-person. Of the 22 respondents, 11 were students, nine were academic researchers, and two were industry researchers. Finally, nine were women, and 13 were men.

Their evaluation of the hybrid event was positive. To the question "Would you attend (in person or online) the 2023 CHIWORK Annual Meeting?" 15 participants said yes, six said maybe, and one said no. On a scale of 1 (worst to 5 (best), 10 participants rated the scientific quality of the meeting as 5, 10 rated it as 4, and two rated it as 3.

The survey also asked responders to tell us about the best and worst aspects of the meeting and suggest improvements. Table 1 shows a selection of the written responses we received.

---

[1] https://www.dagstuhl.de/

Table 1: Sample feedback about best and worst aspects of the meeting, and sample suggested improvements

| |
|---|
| **Best/worst:** |
| "The conversations were a great idea. Good win." |
| "Best - Great job bringing a wonderful community of researchers together. Worst - As an online attendee I wish I could socialize in between sessions in some way." |
| "Best: I really liked the in-person & online chairs. The conversation-style was really good. Worst: Not really worst, but the communication between the chairs could be improved (e.g., through a dedicated channel)" |
| **Suggested improvements** |
| "Enable (serendipitous) social interactions between in-person participants and online ones." |

## 2.2 Town hall

We wrapped up the meeting with a town hall that included both online and in-person participants. We opened a shared file to take down notes and invited all participants in the town hall to help us in this task of taking notes.

Positive comments focused on (1) the efforts to include both online and in-person participants, and (2) the variety of presentation styles. Problems, and suggestions for improvements focused on (1) the difficulties with scheduling for different time zones, (2) improving organizational steps, (3) adding presentation types, and (4) adding tools to bring people closer. Table 1 shows a selection of the comments that were discussed in the town hall meeting.

Table 2: Selection of comments from the town hall

| |
|---|
| **What worked?** |
| "Having two chairs helped to keep track of what's going on." |
| "Conversation style was a lot more engaging in an interview. Was skeptical in the beginning. But feels that this worked, pleasantly surprised." |
| "Having two session chairs to handle both online and in person talks worked really well! The communication was well done and we had plenty of time to arrange everything. (From the chair perspective)." |
| "Liked the variety in the presentation styles and contribution types. Preferred this to separating them out." |
| **What can improve?** |
| "Sending out questions in advance was helpful. But send out earlier." |
| "Room awareness. Sense of people clapping. One room noise was clapping and that was not heard at all." |
| "More room cameras would be good. Constantly capture some sense of the room." |

## 3 LESSON LEARNED

The key lessons we learned from organizing this hybrid event are the following:

1. **Focus on support mechanisms for online participants.** We thought about what problems our online participants might encounter, and we attempted to help them by introducing support mechanisms such as having both an online and in-person chair, spotlighted camera views in Zoom, and a clear audio feed from the in-person venue. The feedback we received indicates that our approach worked. This made the meeting better for everyone, since online participants could more fully participate and thus enrich the meeting.
2. **Online participation is hard. Fixing this might require new approaches.** Our online attendees generally had a good experience. Still, people attending online will continue to face problems from scheduling time away from work and family to attend a presentation, to battling time zones, to having to deal with audio that is sometimes difficult to understand (e.g. due to a poor internet connection, or because an in-person participant does not use a microphone). Some of these problems cannot be solved by technology. However, we can re-think how we bring a community together. For example, the CHIWORK community has weekly 60-minute online conversations which are spread out over several months a year [4]. The short duration of each conversation reduces the pressure of scheduling time away. Also, shorter meetings can more easily be scheduled at times that fit into multiple time zones.

3. **There are many technical improvements to make for hybrid events.** While some problems of bringing together online and in-person audiences cannot be fixed with technology, others certainly can, including (a) online interactions, (b) bringing the sound of applause online, and (c) expanding camera views so that online participants have a good sense of the room.

## ACKNOWLEDGMENTS

We are grateful Susanne Boll, Sarah Fox, Noopur Raval, Max Wilson, Marta Cecchinato, Himanshu Verma, and Marios Constantinides for serving as organizers; to the student volunteers; and especially to Alberta Ansah and Nabil Ch, who were responsible for all logistics at the Annual Meeting. We would also like to thank ACM SIGCHI for supporting CHIWORK through the SIGCHI Development Fund. This work was in part supported in part by the National Science Foundation under grants CMMI-1840085 and CMMI-1840031.

## REFERENCES

[1] Andrew Kun, Orit Shaer, and Shamsi Iqbal. "The Future of Work: COVID-19 and Beyond." IEEE Pervasive Computing 20, no. 04 (2021): 7-8.

[2] Alberta A., Ansah, Adriana S. Vivacqua, Sailin Zhong, Susanne Boll, Marios Constantinides, Himanshu Verma, Abdallah El Ali et al. "Reflecting on Hybrid Events: Learning from a Year of Hybrid Experiences." In *Extended Abstracts of the 2023 CHI Conference on Human Factors in Computing Systems*, pp. 1-4. 2023.

[3] Zoe M. Becerra, Nadia Fereydooni, Andrew L. Kun, Angus McKerral, Andreas Riener, Clemens Schartmüller, Bruce N. Walker, and Philipp Wintersberger. "Interactive workshops in a pandemic: the real benefits of virtual spaces." *IEEE Pervasive Computing* 20, no. 1 (2021): 35-39.

[4] CHIWORK Collective, Naveena Karusala, Nabil Al Nahin Ch, Diana Tosca, Alberta A. Ansah, Emeline Brulé, Nadia Fereydooni et al. "Human-Computer Interaction and the Future of Work." In *CHI Conference on Human Factors in Computing Systems Extended Abstracts*. 2022.